\documentclass{article}

\usepackage{arxiv}

\usepackage[utf8]{inputenc} % allow utf-8 input
\usepackage[T1]{fontenc}    % use 8-bit T1 fonts
\usepackage{hyperref}       % hyperlinks
\usepackage{url}            % simple URL typesetting
\usepackage{booktabs}       % professional-quality tables
\usepackage{amsfonts}       % blackboard math symbols
\usepackage{nicefrac}       % compact symbols for 1/2, etc.
\usepackage{microtype}      % microtypography
\usepackage{lipsum}
\usepackage{graphicx}
\usepackage{hyperref}
\usepackage{amsfonts}
\usepackage{amsmath}
\usepackage{amssymb}
\usepackage{amsthm}
\usepackage{natbib}
\usepackage{mathtools}
\usepackage{float}
\graphicspath{ {./images/} }

\title{A Survey on Algorithmic Developments in Optimal Transport Problem with Applications}

\author{
 Sina Moradi \\
  Etraab Holding\\
Economic Expert\\
Urmia, Iran\\
\texttt{sina\_moradi1993@yahoo.com}\\
}

\newcommand{\vx}{\mathbf{x}}
\newcommand{\vy}{\mathbf{y}}
\newcommand{\vz}{\mathbf{z}}
\newcommand{\vw}{\mathbf{w}}
\newcommand{\vu}{\mathbf{u}}
\newcommand{\vv}{\mathbf{v}}
\newcommand{\va}{\mathbf{a}}
\newcommand{\vb}{\mathbf{b}}
\newcommand{\vone}{\pmb{1}}
\newcommand{\rr}{\mathbb{R}}
\newcommand{\mC}{\mathbf{C}}
\newcommand{\mP}{\mathbf{P}}
\newcommand{\bigO}{\mathcal{O}}
\DeclareMathOperator{\diag}{diag}
\newcommand{\mK}{\mathbf{K}}

\begin{document}
\maketitle
\begin{abstract}
Optimal Transport (OT) has established itself as a robust framework for quantifying differences between distributions, with applications that span fields such as machine learning, data science, and computer vision. This paper offers a detailed examination of the OT problem, beginning with its theoretical foundations, including the classical formulations of Monge and Kantorovich and their extensions to modern computational techniques. It explores cutting-edge algorithms, including Sinkhorn iterations, primal-dual strategies, and reduction-based approaches, emphasizing their efficiency and scalability in addressing high-dimensional problems. The paper also highlights emerging trends, such as integrating OT into machine learning frameworks, the development of novel problem variants, and ongoing theoretical advancements. Applications of OT are presented across a range of domains, with particular attention to its innovative application in time series data analysis via Optimal Transport Warping (OTW), a robust alternative to methods like Dynamic Time Warping. Despite the significant progress made, challenges related to scalability, robustness, and ethical considerations remain, necessitating further research. The paper underscores OT's potential to bridge theoretical depth and practical utility, fostering impactful advancements across diverse disciplines.
\end{abstract}

% keywords can be removed
%\keywords{First keyword \and Second keyword \and More}

\section{Introduction}
\label{sec:intro}
Measuring the difference between two probability distributions is a crucial task underpinning various contemporary data science applications. For example, it is used to evaluate the gap between a model's output distribution and the actual distribution in generative adversarial networks, to analyze intrinsic differences between point clouds in computer graphics, and to examine distributional shifts in transfer learning~\citep{li2023fast}. A widely recognized method for measuring these distances is the optimal transport (OT) problem. Frequently called the earth mover’s or Wasserstein distance, OT has gained substantial attention due to its ability to provide meaningful comparisons between distributions across various contexts. OT theory has become an indispensable tool in research. In economics, it models market equilibrium and resource allocation. In data science and machine learning, it supports techniques like the Wasserstein distance, which is commonly applied in generative models and domain adaptation. OT offers valuable insights into fluid dynamics and population dynamics in physics and biology. The versatility and mathematical rigor of OT make it a vital framework for tackling real-world problems involving efficient resource transfer.

From an algorithmic perspective, the OT problem represents a significant area of research focused on developing computational methods to address OT challenges. This research traces its origins to Monge's pioneering work on mass transportation in the 18th century~\citep{monge1781founding}. OT seeks to determine the most effective method for moving mass distributions or probability across different settings, establishing itself as a crucial concept in mathematics and applied sciences. Its importance has grown substantially in modern machine learning, image processing, and data analysis, spurred by advancements in computational techniques and a deeper understanding of its theoretical foundations~\citep{shen2018wasserstein,wu2019sliced}. The algorithmic development of OT has seen significant progress since Kantorovich introduced a dual formulation in the 1940s, providing a rigorous mathematical framework for the problem~\citep{bouchitte2013monge}. Monge's original formulation focuses on finding a transport map directly transferring mass from the source distribution to the goal while minimizing transportation expenses. However, this approach is notoriously difficult to solve due to its non-convex nature. Kantorovich's relaxation addresses this challenge by introducing a joint transportation plan rather than a direct map, specifying the mass amount moved between each source-target pair. This reformulation transforms the problem into a convex one, making it more computationally tractable.

Recent advancements, including Sinkhorn distances and entropic regularization methods, have significantly improved the efficiency of OT algorithms, enabling them to scale effectively for large datasets~\citep{li2023fast}. Moreover, using Wasserstein distances has enhanced the ability to compare probability distributions, broadening OT's applicability across fields such as economics, machine learning, and beyond~\citep{solomon2015convolutional}. However, several debates persist regarding the computational complexity of OT algorithms, particularly concerning the trade-off between solution accuracy and runtime efficiency. Computing Wasserstein barycenters, an NP-hard problem, remains a major challenge, although approximation methods have demonstrated promising numerical performance~\citep{le2021robust}. Additionally, the ethical considerations surrounding OT's use in areas like fairness in machine learning have prompted discussions about the responsibilities of researchers and practitioners in deploying these methods. The algorithmic perspective on the OT problem highlights its evolving nature and diverse applications, underscoring its importance as a critical area of study for theoretical advancements and real-world solutions.

The OT problem encompasses various applications and is intricately linked with multiple mathematical and computational frameworks. In game theory, OT is utilized to model strategic interactions that involve efficient resource allocation. In information theory, it is a tool for comparing probability distributions with applications in tasks like signal processing and data compression. OT also has significant connections to geometry, particularly in understanding geodesics and Riemannian manifolds, offering a mathematical perspective for solving transportation problems in high-dimensional spaces. Additionally, OT has found applications in healthcare, supporting patient similarity analysis and disease modeling, demonstrating its adaptability and relevance across diverse fields.

This paper is organized as follows: Section~\ref{sec:theory} delves into the theoretical underpinnings of OT, covering its classical formulations and key mathematical properties. Section~\ref{sec:computation} examines computational advancements and innovative algorithmic approaches. Section~\ref{sec:applications} explores the diverse applications of OT across various domains. Lastly, Section~\ref{sec:challenges} outlines emerging trends, outstanding challenges, and potential directions for future research in OT.

\section{Optimal Transport Theory}\label{sec:theory}
\subsection{Theoretical Foundations}
OT theory offers a robust mathematical framework for comparing and transforming probability distributions. Its origins trace back to Monge's early contributions in the eighteenth century, later expanded and formalized by Kantorovich. OT provides a unified approach for addressing both discrete and continuous probability measures. This section introduces the fundamental theoretical concepts of OT; for a comprehensive introduction, readers are encouraged to consult~\citep{peyre2019computational}.

Monge's approach tackles the challenge of identifying a transport map that reduces the expense of transferring mass from one distribution to another. Let $\vx\in\rr^n$ and $\vy\in\rr^m$. The goal of Monge's problem is to identify a mapping $T:\mathcal{X}\to\mathcal{Y}$ that transfers mass from $\alpha$ to $\beta$, where $\mathcal{X}\subseteq\rr^n$ and $\mathcal{Y}\subseteq\rr^m$. The measures $\alpha$ and $\beta$ are defined as: 
\[
\alpha=\sum_{i=1}^{n} \va_i \delta_{x_i}, \quad \beta=\sum_{i=1}^{m} \vb_i \delta_{y_i}
\]
where $\va=[a_1,\dots,a_n]$ and $\vb=[b_1,\dots,b_m]$ represent the weights associated with the locations $\vx\in\mathcal{X}$ and $\vy\in\mathcal{Y}$, respectively. Here, $\delta_z$ denotes the Dirac measure at position $z$, which can be interpreted as a unit of mass concentrated at $z$. When $\va \in \Delta_n$ and $\vb \in \Delta_m$ (the probability simplices in $\mathbb{R}^n$ and $\mathbb{R}^m$, respectively), these measures describe probability distributions. More generally, if all elements of $\va$ and $\vb$ are nonnegative, $\alpha$ and $\beta$ describe positive measures. Monge's problem seeks a mapping $T$ that assigns each $x_i$ in $\alpha$ to a unique $y_j$ in $\beta$ such that the mass of $\beta$ at $y_j$ is equal to the total mass transported from points $x_i$ satisfying $T(x_i) = y_j$, i.e.,
\[
\vb_j = \sum_{i:T(x_i)=y_j} \va_i, \quad \forall j\in\{1,\dots,m\}
\]
This condition can be compactly expressed as $T_\#\alpha = \beta$, in which $T_\#\alpha$ represents the forward mapping of $\alpha$ through $T$. The transport map $T$ is required to minimize a given transportation cost, parameterized by a cost $c(x, y)$, which leads to the optimization problem:
\begin{equation}\label{eq:monge}
    \min_{T} \left\{ \sum_{i} c\left(x_i, T(x_i)\right): T_\#\alpha=\beta \right\}
\end{equation}

Although Monge’s formulation is intuitive and conceptually straightforward, it faces several significant challenges. The problem is inherently non-convex, which makes it difficult to address using conventional optimization techniques. A valid transport map $T$ does not always exist, mainly when the source indicator $\alpha$ is discrete and the target indicator $\beta$ is continuous. Furthermore, Monge’s approach restricts mass splitting at a source point, a limitation that can hinder its applicability in practical scenarios. Despite these difficulties, Monge’s formulation remains highly relevant for applications requiring a direct mapping between source and target distributions. Examples include matching problems in logistics and supply chain management, shape matching and image registration tasks in computer vision, and modeling physical systems governed by strict conservation laws.

Kantorovich’s key contribution lies in relaxing the deterministic nature of transportation, where each source point $x_i$ is assigned to a single target location $y_{\sigma(i)}$ or $T(x_i)$, with $\sigma:\{1,\dots,n\} \to \{1,\dots,m\}$ and $j=\sigma(i)$. Instead of this deterministic approach, Kantorovich allows the mass at a source point $x_i$ to be distributed across multiple target locations. This probabilistic perspective on transport introduces the concept of mass splitting, where mass from a single source can be spread among several targets. To achieve this, Kantorovich replaces the transport map $T$ with a transport plan $\mP \in \mathbb{R}_+^{n \times m}$, which defines a joint probability distribution over the source and target domains. The relaxed problem is expressed as:

\begin{equation}\label{eq:vich}
    \min_{\mP} \left\{ \langle \mC, \mP \rangle : \mP\vone_m = \va, \mP^\top\vone_n = \vb, \mP \in \mathbb{R}_+^{n \times m} \right\},
\end{equation}
where $\langle \mC, \mP \rangle$ represents the total transportation cost, $\mP\vone_m = \va$ ensures that the mass leaving each source matches the distribution $\va$, and $\mP^\top\vone_n = \vb$ ensures that the mass arriving at each target matches the distribution $\vb$.

In contrast to Monge’s formulation, Kantorovich’s relaxation permits the mass at a source point to be distributed among multiple target points. This relaxation makes the problem convex, allowing for the application of efficient optimization techniques. Moreover, a transport plan $\mP$ is guaranteed to exist under mild assumptions, regardless of whether the source and target indicators are discrete or continuous. In this framework, the OT cost establishes a distance metric over the space of probability distributions, known as the Wasserstein distance. This distance is based on a “cost matrix” $\mC$ that quantifies pairwise transport costs and satisfies properties like symmetry, positivity, and triangle inequality. It is a robust metric for comparing distributions, even in high-dimensional settings.

The Kantorovich formulation \eqref{eq:vich} also has an equivalent dual representation that introduces the Kantorovich potentials $\vw \in \mathbb{R}^n$ and $\vz \in \mathbb{R}^m$, which represent the transport costs associated with the source and target domains, respectively. The dual problem is formulated as:
\begin{equation}\label{eq:vichdual}
    \max_{\vw,\vz} \left\{ \langle \vw, \va \rangle + \langle \vz, \vb \rangle : \vw \vone_m^\top + \vone_n \vz^\top \leq \mC \right\}.
\end{equation}

Kantorovich’s relaxation forms the cornerstone of modern optimal transport theory, establishing a rigorous mathematical framework that has paved the way for advancements such as entropic regularization and scalable computational methods.

\subsection{Algorithmic Foundations}\label{sec:3.2}
The OT problem, expressed in the Kantorovich formulations \eqref{eq:vich} and \eqref{eq:vichdual}, can be interpreted as a special instance of LP, with connections to minimum-cost network flow problems. The primal formulation of OT represents a linear program on the feasible set of transportation plans (matrices), minimizing the total transportation cost. The dual formulation introduces Kantorovich potentials, $\vw$ and $\vz$, which are instrumental in understanding the problem's geometric structure and deriving efficient algorithms. The dual problem \eqref{eq:vichdual} possesses a vital property that becomes particularly meaningful in the semidiscrete optimal transport problem. Specifically, the dual variables can be divided based on the constraints of row and column sums, leading to what is known as $\mC$-Transforms.

Considering the dual problem \eqref{eq:vichdual}, fixing $\vw$ (or $\vz$), it can be observed that the optimal vector solution for $\vz$ (or $\vw$) corresponds to the $\mC$-transform of $\vw$ (or $\vz$), denoted by $\vw^{\mC} \in \mathbb{R}^m$ (or $\vz^{\Bar{\mC}} \in \mathbb{R}^n$), defined as:
\[
\left(\vw^{\mC}\right)_j = \min_{i=1,\dots,n} \mC_{ij} - \vw_i, \quad j = 1, \dots, m,
\]
\[
\left(\vz^{\Bar{\mC}}\right)_i = \min_{j=1,\dots,m} \mC_{ij} - \vz_j, \quad i = 1, \dots, n.
\]
The $\mC$-transform $\vw^{\mC}$ (or $\vz^{\Bar{\mC}}$) denotes the maximum vector that complies with the condition stated in \eqref{eq:vichdual}. This implies:
\[
\langle\vw, \va\rangle + \langle\vz, \vb\rangle \leq \langle\vw, \va\rangle + \langle\vw^{\mC}, \vb\rangle,
\]
\[
\langle\vw, \va\rangle + \langle\vz, \vb\rangle \leq \langle\vz^{\Bar{\mC}}, \va\rangle + \langle\vz, \vb\rangle.
\]
These findings reduce the dual formulation \eqref{eq:vichdual} to a maximization problem that is both piecewise affine and concave. Specifically, \eqref{eq:vichdual} becomes equivalent to:
\[
\max_{\vw \in \mathbb{R}^n} \langle\vw, \va\rangle + \langle\vw^{\mC}, \vb\rangle,
\]
\[
\max_{\vz \in \mathbb{R}^m} \langle\vz^{\Bar{\mC}}, \va\rangle + \langle\vz, \vb\rangle.
\]
Starting from an initial $\vw$, it is natural to substitute between $\mC$ and $\Bar{\mC}$ transforms to enhance $\vw$ iteratively. This process leads to the sequence of inequalities:

\begin{align*}
    \langle\vw, \va\rangle + \langle\vw^{\mC}, \vb\rangle &\leq \langle\vw^{\mC\Bar{\mC}}, \va\rangle + \langle\vw^{\mC}, \vb\rangle \\
    &\leq \langle\vw^{\mC\Bar{\mC}}, \va\rangle + \langle\vw^{\mC\Bar{\mC}\mC}, \vb\rangle \\
    &\leq \dots
\end{align*}

Although seeing a clear enhancement in the objective with every iteration may be desirable, alternating $\mC$ and $\Bar{\mC}$ transformations frequently lead to stagnation (i.e., plateau). It has been shown that if $\vw \leq \vw'$, then $\vw^{\mC} \geq \vw'^{\mC}$ and $\vw^{\mC\Bar{\mC}} \geq \vw$. Similarly, $\vz^{\Bar{\mC}\mC} \geq \vz$ and $\vw^{\mC} = \vw^{\mC\Bar{\mC}\mC}$.

Additionally, complementary slackness holds in this formulation. If $\mP^\star$ is an optimal solution to the primal problem \eqref{eq:vich} and $\vw^\star, \vz^\star$ are optimal solutions to the dual problem \eqref{eq:vichdual}, then for all $(i,j)$:
\[
\mP_{ij}^\star \left(\mC_{ij} - \vw_i^\star - \vz_j^\star\right) = 0.
\]
Similarly, if $\mP_{ij}^\star > 0$, then it must hold that $\vw_i^\star + \vz_j^\star = \mC_{ij}$, and if $\vw_i^\star + \vz_j^\star < \mC_{ij}$, then $\mP_{ij}^\star = 0$. Conversely, if the primal \eqref{eq:vich} has a feasible solution $\mP$ and the dual \eqref{eq:vichdual} has complementary feasible solutions $(\vw, \vz)$, then $\mP$ and $(\vw, \vz)$ represent optimal solutions.

\subsection{\texorpdfstring{$\epsilon-$}\ approximation}
An often-used metric for evaluating the numerical performance of approximation algorithms for the OT problem is the computational cost required to compute a solution, \(\Hat{\mP}\), that achieves an \(\epsilon\)-level approximation. This solution satisfies:
\[
\Hat{\mP} \in \left\{ \mP \in \mathbb{R}_+^{n \times m} : \mP \vone_m = \va, \mP^\top \vone_n = \vb \right\},
\]
And ensures:
\[
\langle \mC, \Hat{\mP} \rangle - \langle \mC, \mP^\star \rangle \leq \epsilon,
\]

in which, $\mP^\star$ denotes the optimal solution of the OT problem \eqref{eq:vich}. This inequality is modified to account for the random nature of their outputs for stochastic algorithms and is expressed as:
\[
\mathbb{E} \left[ \langle \mC, \Hat{\mP} \rangle \right] - \langle \mC, \mP^\star \rangle \leq \epsilon.
\]
This formulation allows stochastic methods to measure performance in expectation, maintaining consistency with deterministic approaches while accounting for randomness in their solutions.

\subsection{Entropic Regularization}
The coupling matrix discrete entropy can be shown as:

\[
H(\mP) := -\sum_{i,j} \mP_{ij} \left(\log(\mP_{ij}) - 1\right),
\]

A similar definition applies to vectors. By convention, \(H(\va) = -\infty\) if any entry \(\va_j\) is zero or negative. Since \(\partial^2 H(\mP) = -\diag(1/\mP_{ij})\), combined with the property that \(\mP_{ij} \leq 1\), \(H\) is 1-strongly concave. Entropic regularization in optimal transport involves using $-H$ to estimate solutions to the original transport problem \eqref{eq:vich}. The regularized problem is formulated as:
\begin{equation}\label{eq:hreg}
    \min_{\mP} \left\{\langle \mC, \mP \rangle - \eta H(\mP) : \mP^\top\vone_n = \vb, \mP\vone_m = \va, \mP \in \mathbb{R}_+^{n \times m} \right\}.
\end{equation}

The problem \eqref{eq:hreg} has a unique optimal solution due to the \(\eta\)-strong convexity of the objective function. Introducing an entropic regularization term to the OT problem draws inspiration from concepts in `transportation theory'~\citep{wilson1969use}. As \(\eta \to 0\), it has been shown~\citep{cominetti1994asymptotic} that the solution \(\mP_\eta\) converges to the optimal solution of the Kantorovich problem with the highest entropy among all feasible solutions.
\begin{equation}\label{eq:hregconv1}
    \mP_\eta \to \arg\min_{\mP} \left\{-H(\mP) : \mP \in U(\va, \vb), \langle \mC, \mP \rangle = L_\mC(\va, \vb) \right\},
\end{equation}
where
\[
U(\va, \vb) = \left\{\mP \in \mathbb{R}_+^{n \times m} : \mP^\top\vone_n = \vb, \mP\vone_m = \va \right\},
\]
\[
L_\mC(\va, \vb) = \min_{\mP \in U(\va, \vb)} \langle \mC, \mP \rangle,
\]
and
\[
\min_{\mP \in U(\va, \vb)} \langle \mC, \mP \rangle - \eta H(\mP) \to L_\mC(\va, \vb).
\]

Furthermore, as $\eta$ increases, the solution exhibits a different behavior:
\begin{equation}\label{eq:hregconv2}
    \mP_\eta \to \va \vb^\top.
\end{equation}

Equation \eqref{eq:hregconv1} indicates that for small values of $\eta$, the solution approaches the maximum entropy OT coupling. Conversely, The equation \eqref{eq:hregconv2} indicates that for high $\eta$, the solution approaches the coupling that has the highest entropy between the marginals $\va$ and $\vb$, representing the joint probability distribution of two independent random variables following \(\va\) and \(\vb\). An important point is that as $\eta$ rises, the optimal coupling grows less sparse, resulting in more entries surpassing a specified threshold. Defining the Kullback–Leibler (KL) divergence for couplings as:
\[
KL(\mP \| \mK) = \sum_{i,j} \mP_{ij} \log \left(\frac{\mP_{ij}}{\mK_{ij}}\right) - \mP_{ij} + \mK_{ij},
\]
the unique solution $\mP_\eta$ of \eqref{eq:hreg} can be interpreted as the projection of the Gibbs kernel associated with the cost matrix \(\mC\) onto \(U(\va, \vb)\), where the Gibbs kernel is given by:
\begin{equation}\label{eq:k}
    \mK_{ij} = e^{-\frac{\mC_{ij}}{\eta}}.
\end{equation}
Thus, the solution can be expressed as:
\[
\mP_\eta = \text{Proj}_{U(\va, \vb)}^{KL}(\mK) = \arg\min_{\mP \in U(\va, \vb)} KL(\mP \| \mK).
\]

\subsection{Multimarginal Problems}
Rather than combining two input probability distributions by the Kantorovich formulation \eqref{eq:vich}, it is possible to extend the approach to couple $K$ probability distributions $\{\va_k\}_{k=1}^K$, where $\va_k \in \Delta_{n_k}$. This is accomplished by addressing the ensuing ``multi-marginal optimal transport problem"~\citep{lindheim2023optimal}:
\[
\min_{\mP \in U\left(\{\va_k\}_{k=1}^K\right)} \langle \mC, \mP \rangle,
\]

The set of feasible transport plans is characterized as:

\[
U\left(\{\va_k\}_{k=1}^K\right) = \left\{\mP \in \mathbb{R}_+^{n_1 \times \dots \times n_K} : \va_{k, i_k} = \sum_{l \neq k} \sum_{i_l=1}^{n_l} \mP_{i_1, \dots, i_K}, \ \forall k, \ \forall i_k \right\},
\]
and the transportation cost is given by:
\[
\langle \mC, \mP \rangle = \sum_{k} \sum_{i_k=1}^{n_k} \mC_{i_1, \dots, i_K} \mP_{i_1, \dots, i_K}.
\]

This framework naturally extends the ``entropic regularization" framework \eqref{eq:hreg} to the multi-marginal one:
\[
\min_{\mP \in U\left(\{\va_k\}_{k=1}^K\right)} \langle \mC, \mP \rangle - \eta H(\mP),
\]

where $H(\mP)$ represents the transport plan entropy $\mP$, acting like a regularizer to ensure computational tractability and promote smoother solutions.

\subsection{Unbalanced Optimal Transport}
One significant limitation of traditional OT is the requirement that the two input measures \((\alpha, \beta)\) must have the same total mass. While various workarounds, such as renormalizing the input measures, have been proposed, comprehensive unifying theories have recently been developed to address this issue. One such approach involves the $\phi$-divergence, denoted as $D_\phi$, defined as:
\[
D_\phi(\va \| \vb) = \sum_{i \in \text{Supp}(\vb)} \phi\left(\frac{\va_i}{\vb_i}\right) \vb_i + \phi'_\infty \sum_{i \notin \text{Supp}(\vb)} \va_i,
\]
where $\phi$ is an entropy function, $\text{Supp}(\vb) = \{i \in \{1, \dots, n\} : \vb_i \neq 0\}$, and:
\[
\phi'_\infty = \lim_{x \to \infty} \frac{\phi(x)}{x}.
\]
If $\phi'_\infty = \infty$, it indicates that $\phi$ increases at a rate that surpasses any linear function and is classified as ``superlinear." Every entropy function $\phi$ leads to a related $\phi$-divergence, also called ``Ciszár divergence" or $f$-divergence~\citep{csiszar1967information}.

Based on \cite{liero2018optimal} and \cite{lindheim2023optimal}, the original Kantorovich formulation \eqref{eq:vich} is relaxed to consider arbitrary positive measures $(\va, \vb) \in \mathbb{R}_+^n \times \mathbb{R}_+^m$ by penalizing marginal deviations using a divergence $D_\phi$. This relaxation is related to reducing an optimal transport distance between approximate measures:

\[
\min_{\Tilde{\va}, \Tilde{\vb}} L_\mC(\va, \vb) + \tau_1 D_\phi(\va \| \Tilde{\va}) + \tau_2 D_\phi(\vb \| \Tilde{\vb}) = \min_{\mP \in \mathbb{R}_+^{n \times m}} \langle \mC, \mP \rangle + \tau_1 D_\phi(\mP \vone_m \| \va) + \tau_2 D_\phi(\mP^\top \vone_n \| \vb),
\]
where the parameters $(\tau_1, \tau_2)$ control the balance between penalizing mass variations and transporting the mass. In the limit as $\tau_1 = \tau_2 \to \infty$,  ($\sum_i \va_i = \sum_j \vb_j$ as the ``balanced'' case), this relaxed formulation is the original optimal transport problem alongside hard marginal constraints \eqref{eq:vich}.

\subsection{Wasserstein Barycenters}
``Wasserstein barycenters" extend the notion of a mean or average to the domain of probability distributions within the optimal transport~\citep{lindheim2023optimal}. They represent a ``central'' probability distribution that minimizes the total Wasserstein distance to a given set of input distributions. This concept has become foundational in applications such as clustering, data summarization, and interpolation in high-dimensional probability spaces.

For $K$ probability distributions $\{\va_k\}_{k=1}^{K}$ defined on a shared finite domain $\{x_k\}_{k=1}^{K}$, with associated weights \(w = \{w_1, \dots, w_K\}\) that satisfy \(w_k \geq 0\) for all \(k\) and sum to 1, i.e., \(\sum_k w_k = 1\), the Wasserstein barycenter $\vb$ is the distribution that solves~\citep{lindheim2023optimal}:
\[
\vb \in \arg\min_{\vb \in \Delta_m} \sum_{k=1}^{K} w_k \min_{\mP_k} \left\{ \langle \mC_k, \mP_k \rangle : \mP_k \vone_m = \va_k, \mP_k^\top \vone_{n_k} = \vb, \mP_k \in \mathbb{R}_+^{n_k \times m} \right\}.
\]

The computation of Wasserstein barycenters in their general form is computationally challenging due to the nested optimization over transport plans $\mP_k$ for each distribution and the need to satisfy constraints across all input distributions while determining the barycenter $\vb$. However, the barycenter problem simplifies to a linear program for discrete probability distributions. Specifically, one can seek $K$ couplings $\{\mP_k\}_{k=1}^{K}$ between each input distribution and the barycenter, where the row marginals are shared by construction:
\begin{equation}\label{eq:wb}
    \min_{\vb \in \Delta_m, \{\mP_k \in \mathbb{R}^{n_k \times m}\}_{k=1}^{K}} \left\{ \sum_{k=1}^{K} w_k \langle \mC_k, \mP_k \rangle : \mP_k \vone_m = \va_k, \mP_k^\top \vone_{n_k} = \vb \right\}.
\end{equation}

Although this formulation is a linear program, its computational scale often prohibits general solvers for problems on a medium to large scale. Despite this, Wasserstein barycenters remains a robust framework for analyzing probability distributions and finding applications in machine learning and computational biology. Their theoretical depth and practical relevance make them an indispensable tool in modern data science.

In this section, we have explored the theoretical foundations of the OT problem, including its classical formulations and mathematical properties. Insights from Monge’s and Kantorovich’s formulations, duality principles, and tools like the $\mC$-transform and entropic regularization establish a solid foundation for understanding OT problems. These principles underpin the development of computational methods that make solving OT problems feasible and scalable. For readers interested in advanced topics on OT, we refer to Lindheim's doctoral dissertation~\citep{lindheim2023optimal}.

\section{Algorithms for OT Problem}\label{sec:computation}

OT theory provides a robust mathematical framework for comparing and transforming probability distributions. However, its practical application is often hindered by substantial computational challenges. The core difficulty lies in solving the OT problem efficiently, especially in the context of high-dimensional data or large-scale distributions. Traditional approaches, such as linear programming, yield exact solutions but are computationally intensive and impractical for large datasets. Consequently, notable advancements have been made in developing more efficient algorithms tailored to various problem settings.

Various computational methods have been introduced in recent decades, encompassing both exact solvers and approximate strategies. Exact solvers, such as the simplex and network simplex algorithms, offer precise results but face scalability issues as problem size grows. Approximate methods, including entropic regularization and sliced Wasserstein distances, balance computational efficiency and accuracy, making OT feasible for large-scale applications. Furthermore, specialized algorithms have been designed to handle unique challenges, such as unbalanced measures, and to compute Wasserstein barycenters for tasks like clustering and interpolation.

This section examines the advancements in OT computation, tracing the trajectory from classical exact solvers to contemporary scalable approaches. It highlights key algorithmic innovations, their advantages and limitations, and their influence on real-world applications. By addressing the computational barriers inherent to OT, these methods have significantly broadened their utility across various disciplines, including machine learning, image processing, and economics.

\subsection{Exact Algorithms}
Exact algorithms for the OT problem are fundamental tools designed to compute precise solutions to transportation and allocation tasks. These methods are rooted in classical linear programming and network optimization and offer exact solutions under clearly defined constraints. While their computational demands can be significant, exact algorithms serve as critical benchmarks for evaluating the performance of approximate methods and are indispensable for smaller-scale or high-precision applications. The optimization problem \eqref{eq:vich} is formulated as a linear program and can theoretically be solved in polynomial time using general-purpose algorithms. However, specialized and more efficient algorithms have been developed, which leverage the structure of the OT problem to provide faster and more scalable solutions.

\subsubsection{Algorithms based on Network Flow}
Section~\ref{sec:theory} explains that the OT problem, formulated through the primal \eqref{eq:vich} and dual \eqref{eq:vichdual}, can be interpreted as the minimum-cost network flow problem in linear programs. The network simplex algorithm is well-suited for solving transportation problems because it exploits the underlying network structure~\citep{peyre2019computational}. In the context of OT, this algorithm represents the source and target distributions as nodes in a bipartite graph, with edges weighted by the cost matrix $\mC_{ij}$. It iteratively adjusts the flow along cycles in the network to minimize total transportation cost, maintaining feasibility at every step by satisfying the marginal constraints. The network simplex algorithm has proven to be especially efficient for medium-sized OT problems with sparse cost matrices. Notably, \cite{doi:10.1287/opre.41.2.338} provided the first proof of the algorithm's polynomial time complexity. Subsequently, \cite{tarjan1997dynamic} improved the complexity bound to $\bigO ((n + m)nm \log(n + m) \log ((n + m)\Vert C \Vert_{\infty}) )$ by introducing more efficient data structures for selecting pivoting edges. This enhancement significantly optimized the algorithm's performance, making it more practical for problems with large but structured cost matrices.

\subsubsection{Algorithms based on Combinatorial Optimization}
The auction algorithm, introduced by Bertsekas~\citep{bertsekas1981new,bertsekas1988dual,bertsekas1992auction}, is a combinatorial optimization method tailored for assignment problems, which represent a special case of the OT problem. The algorithm iteratively modifies $(S, \xi, \vz)$, where $S \subseteq \{1, \dots, n\}$ is a set of assigned points, $\xi$ is a partial assignment vector mapping elements of $S$ injectively to $\{1, \dots, m\}$, and $\vz$ is a dual vector. 

Given a source index $i$, the auction algorithm evaluates not only the optimal assignment but also the second-best option, using the $\mC$-transform:
\[
j_i^1 \in \arg\min_{j \in \{1, \dots, m\}} \mC_{ij} - \vz_j, \quad 
j_i^2 \in \arg\min_{j \in \{1, \dots, m\} \setminus j_i^1} \mC_{ij} - \vz_j.
\]
Based on these indices, the algorithm updates $\vz$, $S$, and $\xi$ as follows:
\begin{enumerate}
    \item Update the dual vector $\vz$:
    \begin{align*}
        \vz_{j_i^1} &\leftarrow \vz_{j_i^1} - \left(\left(\mC_{i,j_i^2} - \vz_{j_i^2}\right) - \left(\mC_{i,j_i^1} - \vz_{j_i^1}\right) + \epsilon\right) \\
        &= \mC_{i,j_i^1} - \left(\mC_{i,j_i^2} - \vz_{j_i^2}\right) - \epsilon.
    \end{align*}
    \item Update $S$ and $\xi$: If there exists $i' \in S$ such that $\xi_{i'} = j_i^1$, remove $i'$ from $S$ by setting $S \leftarrow S \setminus \{i'\}$, and add $i$ to $S$ with $S \leftarrow S \cup \{i\}$.
\end{enumerate}

The algorithm starts with $S = \emptyset$, $\xi = \emptyset$, and $\vz = \pmb{0}_m$, and iteratively applies the steps above until all points are assigned, i.e., $S = \{1, \dots, n\}$. With at most $N = n \|\mC\|_\infty / \epsilon$ steps, the resulting assignment incurs a cost that is at most $n\epsilon$ suboptimal.

Building on the auction algorithm, \cite{lahn2019graph} proposed a deterministic primal-dual approach for solving the discrete OT problem with $\epsilon$-approximation guarantees. By adapting the scaling algorithm of Gabow and Tarjan, this method achieves a computational complexity of:
\[
\mathcal{O}\left(\frac{n^2}{\epsilon} + \frac{n}{\epsilon^2}\right).
\]
This approach involves scaling demands and supplies from a graph-theoretic perspective, followed by iterative updates of dual weights and adjustments to the transport plan along admissible paths. Unlike iterative methods like Sinkhorn’s algorithm, the primal-dual approach is numerically stable, even for small $\epsilon$ values, and avoids logarithmic dependencies in runtime.

Exact algorithms, including the network simplex and auction methods, form the foundation of optimal transport theory. While their computational intensity limits scalability for large-scale problems, they are invaluable for small- to medium-sized applications and serve as benchmarks for evaluating approximate methods. These advancements ensure that exact algorithms remain relevant across diverse applications and inspire new computational strategies.

\subsection{Iterative or Approximate Algorithms}

\subsubsection{Sinkhorn-based Algorithms}
Entropic regularization is used by Sinkhorn's algorithm to iteratively approximate optimal transport solutions~\citep{sinkhorn1964relationship}. This algorithm efficiently solves the regularized problem \eqref{eq:hreg}, making it particularly suitable for large-scale datasets and high-dimensional applications. The solution to \eqref{eq:hreg} is unique and can be expressed as:
\[
\mP_{ij} = \vu_i \mK_{ij} \vv_j,
\]
where $\mK$ is defined in \eqref{eq:k}, and $\vu, \vv \in \mathbb{R}_+^{n \times m}$ are scaling variables. This factorization is compactly written as $\mP = \diag(\vu) \mK \diag(\vv)$, where 
\[
\diag(\vu) \mK \diag(\vv) \vone_m = \va, \quad \diag(\vv) \mK^\top \diag(\vu) \vone_n = \vb.
\]
or equivalently:
\[
(\mK \vv) \odot \vu = \va, \quad (\mK^\top \vu) \odot \vv = \vb,
\]
where $\odot$ denotes element-wise multiplication. To solve these equations, the algorithm iteratively updates $\vu$ and $\vv$ as follows:
\begin{enumerate}
    \item Initialize $\vu = \vone_n$ and $\vv = \vone_m$,
    \item Iteratively update:
    \[
    \vu^{(l+1)} \leftarrow \frac{\va}{\mK \vv^{(l)}}, \quad \vv^{(l+1)} \leftarrow \frac{\vb}{\mK^\top \vu^{(l+1)}},
    \]
    \item Recover the transport plan using $\mP = \diag(\vu) \mK \diag(\vv)$.
\end{enumerate}
The algorithm converges to the regularized optimal solution, with each iteration requiring $\mathcal{O}(mn)$ operations.

For the case $n = m$, \cite{altschuler2017near} introduced an extension to the Sinkhorn algorithm by reanalyzing its iterations and proposing a new greedy coordinate descent algorithm called Greenkhorn. The traditional Sinkhorn method iteratively adjusts the rows and columns of \(\mK\) to ensure that the marginal constraints are met. Altschuler et al. showed that the number of iterations for Sinkhorn is bounded by $\mathcal{O}(\log(s/l) / \epsilon^2)$, where $s = \sum_{i,j} \mK_{ij}$ and $l = \min_{i,j} \mK_{ij}$. Greenkhorn achieves the same convergence guarantee but improves efficiency in practice by updating only one row or column of $\mK$ at a time, rather than all rows and columns simultaneously.

Greenkhorn identifies the best row or column to update greedily by maximizing the function $\rho: \mathbb{R}_+ \times \mathbb{R}_+ \to [0, +\infty]$:
\[
\rho(a, b) = b - a + a \log \frac{a}{b}.
\]
The row and column indices for each iteration are selected as:
\[
i \leftarrow \arg\max_i \rho(\va_i, (\mK \vv)_i), \quad
j \leftarrow \arg\max_j \rho(\vb_j, (\mK^\top \vu)_j).
\]
Greenkhorn performs an exact line search for the chosen coordinate, leveraging a simple closed-form solution. This reduces computational complexity per iteration, requiring updates to only $\mathcal{O}(n)$ entries of $\mK$ instead of $\mathcal{O}(n^2)$. The total complexity of Greenkhorn is $\mathcal{O}(n^2 / \epsilon^3)$, making it particularly advantageous for sparse datasets.

Both Sinkhorn and Greenkhorn algorithms have expanded the practical applicability of OT by enabling scalable computations while maintaining theoretical guarantees.

\subsubsection{Primal-Dual Algorithms}
\cite{dvurechensky2018computational} proposed an algorithm for addressing OT problems under different regularization schemes. They also assumed $m=n$. They demonstrated that the Sinkhorn's iteration bound~\citep{altschuler2017near} represents the worst-case estimate. They established that the iteration count for the Sinkhorn algorithm is upper-bounded by $$\bigO\left(\frac{n^2 \ln n\Vert\mC\Vert_{\infty}}{\epsilon^2}\right).$$
Given linear operator $\mathbf{A}:E\to H$ with $E \text{ and }H$ being some finite real vector space and $E$ being convex with some chosen norm $\Vert\cdot\Vert_E$, a given $\mathbf{r} \in H$, and $f(\vx)$ being a $\gamma-$strongly convex on $E$, they analyze a problem
\begin{equation}\label{eq:genp}
    \min_{\vx\in E} \{f(\vx): \mathbf{A}\vx=\mathbf{r}\}
\end{equation}
Its Lagrange dual problem is
\begin{equation}\label{eq:gend}
    \min_{\pmb{\lambda}\in H^*} \left\{\langle\pmb{\lambda}, \mathbf{r}\rangle + \max_{\vx\in E}(-f(x) - \langle \mathbf{A}^\top\pmb{\lambda},\vx\rangle)\right\}
\end{equation}
and $\nabla\phi(\pmb{\lambda})$ is $L-$Lipschitz-continuous with $L\leq \Vert\mathbf{A}\Vert_{E\to H}^2/\gamma$. It is assumed that the dual problem has a solution and there exists some $R > 0$ where $\Vert\pmb{\lambda}^*\Vert_2 \leq R < + \infty$ and $\pmb{\lambda}^*$ is the optimal dual solution with objective value of $\Vert\pmb{\lambda}^*\Vert_2$. They proved the convergence rate of their algorithm, called Adaptive Primal-Dual Accelerated Gradient Descent (APDAGD), is
\[
\Vert\mathbf{A}\Hat{\vx}_k - \mathbf{r} \Vert_2 \leq \bigO\left(\frac{1}{k^2}\right)
\]
\[
\Vert\Hat{\vx}_k - \vx^*\Vert_{E} \leq \bigO\left(\frac{1}{k}\right)
\]
\[
f(\Hat{\vx}_k) - f^* \leq \bigO\left(\frac{1}{k^2}\right)
\]

This algorithm extends the accelerated mirror descent framework into a primal-dual approach. Unlike traditional methods, it introduces a linesearch strategy and employs an online stopping rule based exclusively on the duality gap and the degree of constraint violation.

The OT problem \eqref{eq:hreg} can be seen as a specific instance of this general optimization framework by setting $\vx=\text{Vec}(\mP)\in\rr^{n^2}$ and
\[
f(\vx)= \langle \mC,\mathbf{X}\rangle - \gamma H(\mathbf{X})
\]
\[
\mathbf{r} = [\va^\top,\vb^\top]
\]
\[
(\mathbf{A}\vx)^\top = \left[(\mP\vone)^\top, (\mP^\top\vone)^\top\right]
\]
Then, based on APDAGD, they proved the complexity bound $\bigO(n^{2.5}/\epsilon)$ arithmetic operations.

\cite{jambulapati2019direct} proposed an algorithm leveraging dual extrapolation. This parallelizable method, that is an innovative primal-dual first-order, achieves additive $\epsilon-$accuracy with near-linear time complexity, offering an advancement over earlier approaches that depended on matrix scaling or second-order Newton iterations. The approach leverages area-convexity to simplify the analysis and improve runtime guarantees. Their method builds significantly on advancements in solving the maximum flow problem and, more generally, on two-player games constrained within a simplex ($\ell_1$ ball) and a box ($\ell_{\infty}$ ball). A minimax game between a simplex and a box is formulated for \eqref{eq:hreg}. $\mathbf{d}$ was defined as $\text{Vec}(\mC)$, $\mathbf{r}$ as $ [\va^\top,\vb^\top]$, and $\mathbf{A}\in\{0,1\}^{2n\times m}$ is a \(0\)-\(1\) matrix on \(V \times E\) where \(\mathbf{A}_{ve} = 1\) if and only if \(v\) is an endpoint of edge \(e\). Then, they showed that problem \eqref{eq:hreg} is equivalent to $\ell_1$ regression objective over the simplex
\begin{equation}\label{eq:bad}
    \min_{\vx\in\Delta_m} \mathbf{d}^\top \vx + 2 \Vert \mathbf{A}\vx - \mathbf{r}\Vert_1 \Vert\mathbf{d}\Vert_{\infty} 
\end{equation}
By showing that this problem can be reformulated through a primal-dual perspective as  
\[
\min_{\vx \in \Delta_m} \max_{\vy \in [-1,1]^{2n}} \mathbf{d}^\top\vx + 2 \left(\vy^\top\mathbf{A}\vx - \mathbf{r}^\top\vy\right) \Vert\mathbf{d}\Vert_{\infty},
\]
they established the existence of an algorithm that, given input \(\epsilon\), achieves parallel depth of \(\bigO\left(\frac{ \log\gamma \log n\Vert\mathbf{d}\Vert_{\infty}}{\epsilon}\right)\), where \(\gamma = \frac{\log n \cdot \Vert\mathbf{d}\Vert_{\infty}}{\epsilon}\), and complexity of \(\bigO\left(\frac{n^2 \log\gamma\log n  \Vert\mathbf{d}\Vert_{\infty}}{\epsilon}\right)\), producing \(\Tilde{\vx}\), an \(\epsilon\)-additive approximation to the objective in \eqref{eq:bad}. Further, they use area-convex regularization $r$ to ensure convergence and simplify theoretical guarantees without requiring strong convexity in traditional norms.

The paper outlines a dual extrapolation algorithm summarized as:
\begin{enumerate}
    \item Initialization: Begin with $s_0=0$ and $z_0-0$ as the minimizer of the regularizer $r$.
    \item Compute $z_t$ using the proximal operator
    \[
    z_t=\text{Prox}_r(s_t)
    \]
    where \(\text{Prox}_r(s_t)\) finds the point that minimizes the Bregman divergence relative to \(r\).
    \item Update the auxiliary variable $w_t$
    \[
    w_t=\text{Prox}_r\left(s_t + \frac{1}{\kappa}g(z_t)\right)
    \]
    where $g$ represents the gradient operator.
    \item Update the dual iterate $s_{t+1}$
    \[
    s_{t+1}=s_t + \frac{1}{\kappa}g(w_t)
    \]
\end{enumerate}
They introduced a rounding step to transform approximate solutions into feasible transport plans without degrading solution quality. Given an approximate transport plan $\Tilde{X}$, a rounding step adjusts it to satisfy marginal constraints $\va$ and $\vb$ while preserving accuracy. Finally, this parallelizable primal-dual first-order algorithm based on dual extrapolation proves the complexity bound $\bigO(n^2/\epsilon)$ arithmetic operations.

\cite{guo2020fast} proposed two algorithms that aim to compute entropic-regularized OT distances with theoretical guarantees and better scalability than existing methods like Sinkhorn iterations or deterministic primal-dual gradient descent methods. These algorithms, called Accelerated Primal-Dual Randomized Coordinate Descent (APDRCD) and Accelerated Primal-Dual Greedy Coordinate Descent (APDGCD), are built upon the general framework of accelerated primal-dual coordinate descent. Assuming $m=n$, they formulated the dual problem for the entropic regularized OT problem as
\[
\min_{\alpha,\beta\in\rr^n} \phi(\alpha,\beta):= \eta \sum_{i=1}^{n} \sum_{j=1}^{n} \exp{\left(-\frac{\mC_{ij} - \alpha_i - \beta_j}{\eta} - 1\right)} - \langle\beta,\vb\rangle - \langle\alpha,\va\rangle
\]
where $\phi(\alpha,\beta)$ is smooth, convex, and satisfies key properties like Lipschitz continuity.

The APDRCD algorithm consists of two primary components. Initially, it leverages the convexity of the dual objective function by executing a randomized accelerated coordinate descent step as a subroutine. As the second phase, a weighted average of the previous iterations is calculated to derive an approximate solution to the primal problem using the approximate solutions obtained for the dual problem. The APDRCD algorithm begins with the auxiliary sequence $\{\theta_i\}$ and two auxiliary dual variable sequences, $\{\lambda_i\}$ and $\{z_i\}$. The sequence $\{\theta_i\}$ is critical in averaging. In contrast, the dual variable sequences execute the accelerated randomized coordinate descent on the dual objective function $\phi$ as a subroutine. In other words, in each iteration $k$,
\begin{enumerate}
    \item Set $y_k=(1-\theta_k)\lambda^k+\theta_k z^k$
    \item Compute $\vx_k = \frac{1}{\mC_k}\left(\sum_{j=1}^{k}\frac{\vx(y_j)}{\theta_j}\right)$.
    \item Randomly sample one coordinate $i_k$ where $i_k\in \{1, 2, ..., 2n\}$
    \item Update $\lambda_{i_k}^{k+1}=y_{i_k}^k - \frac{1}{L}\nabla_{i_k}\phi(y^k)$.
    \item Update $z_{i_k}^{k+1}=z_{i_k}^k - \frac{1}{2nL\theta_{i_k}}\nabla_{i_k}\phi(y^k)$
    \item Update $\mC_k=\mC_k + \frac{1}{\theta_k}$
\end{enumerate}
It was demonstrated that the APDRCD algorithm for approximating optimal transport achieves a solution with a total of
\[
\bigO\left(\frac{n^{2.5} \sqrt{\log n \Vert\mC\Vert_{\infty}}}{\epsilon}\right)
\]
arithmetic operations. They introduced a greedy variant of the APDRCD algorithm named APDGCD. This algorithm, like APDRCD, consists of two main components. In the first step, instead of employing randomized accelerated coordinate descent on the dual objective function, APDGCD selects the coordinate with the largest absolute gradient value of the dual objective function for \eqref{eq:hreg}. The second step uses the key averaging technique from APDRCD, computing a weighted average over previous iterations to derive an approximate solution for the primal problem based on the approximate dual solutions. In other words, in step 3 above,
\[
i_k = \arg\max_{i\in\{1,\dots,2n\}} \left|\nabla_{i_k}\phi(y^k)\right|
\]
Finally, they proved that both algorithms guarantee $\epsilon-$approximation of the OT distance in $\bigO(n^{2.5}/\epsilon)$ operations, independent of problem sparsity.

\cite{lin2022efficiency} considered the entropic regularized formulation \eqref{eq:hreg}. They improved the analysis of Greenkhorn, showing it attains a computational complexity of $\bigO(n^2/\epsilon^2)$ (best-known bound for Sinkhorn but requires fewer updates per iteration). They introduced an algorithm that extends APDAGD~\citep{dvurechensky2018computational} by incorporating a predefined mirror mapping \(\phi\). They proved that their algorithm, called adaptive primal-dual accelerated mirror descent (APDAMD), achieves a \(\bigO(n^2\sqrt{\delta}/\epsilon)\) computational complexity, where \(\delta > 0\) represents the regularity of \(\phi\) for the \(\ell_\infty\) norm. They demonstrated the complexity bound of \(\bigO(\min\{n^{9/4}/\epsilon, n^2/\epsilon^2\})\) for APDAGD~\citep{dvurechensky2018computational} is incorrect and provided a refined complexity of \(\bigO(n^{2.5}/\epsilon)\), which is less favorable with respect to \(n\). Additionally, they proposed a variant of Sinkhorn based on an estimated sequence and proved its complexity of \(\bigO(n^{7/3}/\epsilon^{4/3})\) that is deterministic and accelerated. Notably, accelerated Sinkhorn employs exact minimization for the primary iterates alongside a secondary sequence utilizing coordinate gradient updates and monotone search. They further demonstrated that this variant of Sinkhorn surpasses Sinkhorn and Greenkhorn with \(1/\epsilon\) and outperforms APDAGD in terms of \(n\). Assuming $m=n$, they formulated the dual problem for the entropic regularized OT problem as
\[
\min_{\alpha,\beta\in\rr^n} \phi(\alpha,\beta):= \eta \log\left( \sum_{i=1}^{n} \sum_{j=1}^{n} \exp{\left(-\frac{\mC_{ij} - \alpha_i - \beta_j}{\eta}\right)}\right) - \langle\beta,\vb\rangle - \langle\alpha,\va\rangle
\]
By substituting \(\vu = \frac{\alpha}{\eta}\) and \(\vv = \frac{\beta}{\eta}\), the problem can be reformulated as  
\[
\min_{\vu, \vv \in \rr^n} \phi(\vu, \vv) := \log\left(\Vert B(\vu, \vv)\Vert_1\right) - \langle \vv, \vb \rangle - \langle \vu, \va \rangle,
\]
where  
\[
B(\vu, \vv) = [B_{ij}] \in \rr^{n \times n}, \quad B_{ij} = e^{\vu_i + \vv_j - \frac{\mC_{ij}}{\eta}}.
\]
For APDAMD, they considered the same general optimization problem as~\citep{dvurechensky2018computational} with primal \eqref{eq:genp} and dual \eqref{eq:gend}. The main steps of this algorithm in updating primal and dual variables are
\begin{enumerate}
    \item Update average step by $\mu_{k+1}=\frac{\alpha_{k+1}z_k+\Bar{\alpha}_k\lambda_k}{\Bar{\alpha}_{k+1}}$.
    \item Update mirror descent by $z_{k+1}=\arg\min_z\left\{(z-\mu_{k+1})^\top\nabla\phi(\mu_{k+1}) + B_\phi(z,z_k)/\alpha_{k+1}\right\}$.
    \item Update dual by $\lambda_{k+1}=\frac{\alpha_{k+1}z_{k+1}+\Bar{\alpha}_k\lambda_k}{\Bar{\alpha}_{k+1}}$
\end{enumerate}
APDAMD achieves a computational complexity of \(\bigO(n^2\sqrt{\delta}/\epsilon)\), with \(\delta > 0\) representing the regularity of the mirror map. It outperforms traditional methods like Sinkhorn, offering superior theoretical guarantees and practical effectiveness. 

The Accelerated Sinkhorn algorithm enhances the computational efficiency of solving the entropic regularized optimal transport problem, especially for large-scale applications. Building on the classical Sinkhorn method, this approach integrates Nesterov's acceleration to achieve faster convergence while maintaining scalability. The Accelerated Sinkhorn algorithm uses iterative updates for the dual variables $\vu_k \text{ and } \vv_k$
\[
\vu_{k+1} = \vu_k - \frac{1}{2\theta_k} \left(B(\vu_k,\vv_k)\vone - \va\right)
\]
\[
\vv_{k+1} = \vv_k - \frac{1}{2\theta_k} \left(B(\vu_k,\vv_k)^\top\vone - \vb\right)
\]
where $\theta_k$ is updated using Nesterov’s acceleration. The algorithm attains a computational complexity of \(\bigO(n^{7/3}/\epsilon^{4/3})\), representing a substantial enhancement compared to the traditional Sinkhorn method's \(\bigO(n^2/\epsilon^2)\). Experimental results demonstrate its effectiveness in reducing runtime and iteration count, making it a robust choice for large-scale OT problems.

A novel method utilizing Nesterov's smoothing technique was proposed by \cite{an2022efficient} to enhance computational efficiency as well as the accuracy of solving the OT problem. The algorithm offers theoretical and practical advantages over existing solutions by focusing on the dual Kantorovich problem without entropic regularization in \eqref{eq:vich} (for general measure) and leveraging advanced optimization methods. Remarkably, their main contribution is an Accelerated Gradient Descent method for the dual Kantorovich problem. They applied Nesterov's smoothing technique to replace the non-smooth Kantorovich functional with a smooth Log-Sum-Exp approximation. This reformulation transforms the dual problem into a smooth, unconstrained optimization problem, enabling efficient resolution through the Fast Iterative Shrinkage-Thresholding Algorithm (FISTA).

Using the $\mC-$transforms discussed in Section~\ref{sec:3.2}, they defined an equivalent formulation for \eqref{eq:vich}. Nesterov's smoothing approximates the \(\mC\)-transforms with the Log-Sum-Exp function, resulting in a smooth Kantorovich functional. FISTA was then applied to the smoothed Kantorovich functional to solve the optimal transport problem. The Accelerated Gradient Descent with Nesterov’s Smoothing algorithm offers an efficient solution for OT problems, addressing the computational challenges of existing methods like Sinkhorn. The algorithm achieves a complexity of \(\bigO(n^{2.5}\log n / \epsilon)\), offering better performance than Sinkhorn in high-precision tasks and ensuring theoretical guarantees for both convergence and approximation accuracy. 

\cite{chambolle2022accelerated} focused on Hybrid Primal-Dual (HPD) methods applied to solving OT for both with and without entropic regularization and Wasserstein Barycenter problems, introducing significant improvements to existing algorithms. They extended the analysis of HPD methods with line search to settings where the dual space is equipped with a Bregman divergence, achieving accelerated convergence rates $\bigO(1/n^2)$ under strong convexity assumptions. Also, they proposed a new scaled entropy kernel to improve numerical stability, avoiding issues related to logarithms of small values and yielding sparse solutions desirable in specific applications. Within the HPD framework, the following general optimization problem was analyzed:  
\begin{equation}\label{eq:hpd}
    \min_{\vx \in X}\max_{\vy \in Y} g(\vx) - h^*(\vy) + \langle \mK\vx, \vy \rangle,
\end{equation}  
with convex sets \(X\) and \(Y\) belonging to the normed spaces \((\mathcal{X}, \Vert\cdot\Vert_{\mathcal{X}})\) (referred to as the primal space) and \((\mathcal{Y}, \Vert\cdot\Vert_{\mathcal{Y}})\) (referred to as the dual space). The linear operator \(\mathbf{K}: \mathcal{X} \to \mathcal{Y}^*\) maps between these spaces, and the functions \(g\) and \(h^*\) are proper, convex, and lower semicontinuous, defined on \(X\) and \(Y\), respectively. Both are equipped with \(
D(z, \Bar{z}) = \xi(z) - \xi(\Bar{z}) - \langle \nabla \xi(\Bar{z}), z - \Bar{z} \rangle,
\) that is Bregman divergence with \(\xi\) is a 1-strongly convex prox-function relative to the norm of the space. The function \(\xi\) is differentiable within \(\text{int}(\text{dom} \, \xi)\) and satisfies the property that \(\Vert \nabla \xi(x)\Vert \to \infty\) as \(x\) approaches \(\partial (\text{dom} \, \xi)\). Starting with the initial points \(\vx^0, \Bar{\vx}^0 \in \text{dom} \, \xi_{\mathcal{X}}\) and \(\vy^0 \in \text{dom} \, \xi_{\mathcal{Y}}\), along with predefined nonnegative sequences \(\{\tau_k\}_k\), \(\{\sigma_k\}_k\), and \(\{\theta_k\}_k\), the HPD method with line search proceeds as follows for its main iteration:
\begin{enumerate}
    \item $\tau_k \leftarrow \rho\tau_k$, $\theta_k \leftarrow \tau_k/\tau_{k-1}$, $\sigma_k \leftarrow  \beta_k\tau_k$.
    \item $\Bar{\vx}^k \leftarrow  \theta^k(\vx^k - \vx^{k-1}) + \vx^k$.
    \item $\vy^{k+1} \leftarrow  \arg \min_{y\in \mathcal{Y}}  h^* (\vy)  + \frac{D_{\mathcal{Y}}(\vy, \vy^k)}{\sigma_k} - \langle \mathbf{K} \Bar{\vx}^k, \vy\rangle$.
    \item $\vx^{k+1} \leftarrow  \arg \min_{x\in \mathcal{X}}  g(\vx) + \langle \mathbf{K}\vx, \vy^{k+1}\rangle + \frac{\Vert \vx - \vx^k\Vert_2^2}{2\tau_k}$
\end{enumerate}
Then, they formulated \eqref{eq:vich} in the form of \eqref{eq:hpd} as
\[
\min_{\mP\in \Delta}\max_{\vu,\vv} \langle \mC,\mP\rangle + \langle\vu,\va - \mP\vone_n\rangle + \langle\vv,\vb - \mP^\top\vone_n\rangle
\]
where $\Delta=\{\mP\in\rr_+^{n\times n}: \vone^\top\mP\vone_n=1\}$; and entropic regularized OT in \eqref{eq:hreg} in the form of \eqref{eq:hpd} as
\[
\min_{\mP\in \Delta}\max_{\vu,\vv} \langle \mC + \eta \ln\mP,\mP\rangle + \langle\vu,\va - \mP\vone_n\rangle + \langle\vv,\vb - \mP^\top\vone_n\rangle
\]
and, the Wasserstein Barycenter problem \eqref{eq:wb} n the form of \eqref{eq:hpd} as

\[
\min_{\vb\in\Delta,\mP_k\in \Delta}\max_{\vu,\vv} \sum_{k=1}^{K} w_k \left( \langle \mC_k,\mP_k\rangle + \langle\vu_k,\va_k - \mP_k\vone_n\rangle + \langle\vv_k,\vb - \mP_k^\top\vone_n\rangle \right)
\]
Building on the HPD framework, they derived a complexity of \(\bigO(\Vert\mC\Vert n^{2.5} / \epsilon)\) for the OT problem and \(\bigO(K \Vert\mC\Vert n^{2.5} / \epsilon)\) for the Wasserstein barycenter problem, both in terms of arithmetic operations.

The algorithm, developed by \cite{xie2022accelerated}, offers significant improvements in solving optimization problems with a complexity of \(\bigO(n^{2.5}/\epsilon)\). This algorithm, called Primal-Dual Accelerated Stochastic Gradient Descent (PDASGD), leverages advanced techniques such as variance-reduction in stochastic gradients to speed up convergence and incorporates Nesterov’s acceleration for improved efficiency. Additionally, the algorithm operates in two stages: an approximate solution is first computed for the entropy-regularized OT problem, followed by a refinement step to ensure the constraints are met. They examined the same general optimization framework as presented in~\citep{dvurechensky2018computational,lin2022efficiency}, with the Lagrangian function defined as:
\begin{align*}
    \phi(\pmb{\lambda}) &= -\langle\pmb{\lambda}, \mathbf{r}\rangle + \max_{\vx\in \rr^q}(-f(\vx) + \langle \mathbf{A}^\top\pmb{\lambda},\vx\rangle) \\
    &= -\langle\pmb{\lambda}, \mathbf{r}\rangle -f(\vx(\pmb{\lambda})) + \langle \mathbf{A}^\top\pmb{\lambda},\vx(\pmb{\lambda})\rangle
\end{align*}
and $\vx(\pmb{\lambda})=\arg\max_{\vx\in \rr^q}(\langle \mathbf{A}^\top\pmb{\lambda},\vx\rangle -f(\vx))$. They assumed that
\begin{itemize}
    \item $\phi$ is separable and can be expressed as:
    \[
    \phi(\pmb{\lambda}) = \frac{1}{k} \sum_{i=1}^{k} \phi_i(\pmb{\lambda})
    \]
    \item The function \(\phi_i(\pmb{\lambda})\) exhibits convexity and \(L_i\)-smoothness with respect to the \(\Vert\cdot\Vert_2\) norm. The average smoothness parameter is represented as:
    \[
    \Bar{L} = \frac{1}{k} \sum_{i=1}^{k} L_i.
    \]
    \item The relationship between \(\vx(\pmb{\lambda})\) and the derivative or subgradient of \(\phi(\pmb{\lambda})\) is given by the following equation:
    \[
    \nabla \phi(\pmb{\lambda}) = \mathbf{A} \vx(\pmb{\lambda}) - \mathbf{r}
    \]
    \item The primal objective function \(f(\vx)\) is assumed to exhibit strong convexity concerning its associated norm. 
\end{itemize}
PDASGD iterations include two main loops:
\begin{itemize}
    \item Inner Loop: Updates dual variables using variance-reduced gradients and Katyusha momentum. In each iteration $t$, 
    \[
    \tau^1_{t} = \frac{2}{4+s};\quad \gamma_t = \frac{1}{9\Bar{L}\tau^1_{t}}, \quad u^t = \nabla\phi(\Tilde{\lambda}^t)
    \]
    for \(j\) ranges from 0 to \(m-1\),
    \begin{enumerate}
        \item $k\leftarrow j+tm$
        \item $\lambda_{k+1}\leftarrow \tau_2 \Tilde{\lambda}^t + \tau^1_{t} z_k  + (1 - \tau_2 - \tau^1_{t}) y_k$
        \item $\Tilde{\nabla}_{k+1} \leftarrow u^t + \frac{\nabla_{\phi_i}(\lambda_{k+1}) - \nabla_{\phi_i}(\Tilde{\lambda}^t)}{h p_i}$ where $i\sim\text{Categorical}(p_1,\dots,p_h)$, \(i\in\{1, 2, \dots, h\}\), and \(p_i = \frac{L_i}{h \Bar{L}}\).
        \item $z_{k+1} \leftarrow z_k - \frac{\gamma_t }{2}\Tilde{\nabla}_{k+1}$
        \item $y_{k+1} \leftarrow \lambda_{k+1} - \frac{1}{9\Bar{L}} \Tilde{\nabla}_{k+1}$
    \end{enumerate}
    
    \item Outer Loop: Aggregates inner-loop results to ensure feasibility and convergence. For $t=0,\dots,T-1$,
    \begin{enumerate}
        \item $\Tilde{\lambda}^{t+1} \leftarrow \frac{1}{m} \sum_{j=1}^{m} y_{tm+j}$
        \item $D_t \leftarrow D_t + \vx(\Hat{\lambda}_t)/\tau^1_{t}$, where $\Hat{\lambda}_t$ is selected randomly from $\{\lambda_{tm+1}, \dots , \lambda_{tm+m}\}$
        \item $\mC_t \leftarrow \mC_t + 1/\tau^1_{t}$, $\vx^t = D_t/C_t$
    \end{enumerate}
\end{itemize}
They proved the convergence rate of PDASGD if applied to problem \eqref{eq:vich} and \eqref{eq:hreg},
\[
f(\vx^k)-f(\vx^\star) \leq \bigO\left(\frac{1}{k^2}\right)
\]
\[
\Vert\mathbb{E}\left[\mathbf{A}\vx^k - \mathbf{r}\right] \Vert_2 \leq \bigO\left(\frac{1}{k^2}\right)
\]
The algorithm's complexity is $\bigO(n^{2.5}/\epsilon)$, making it significantly faster than previous methods, particularly for large-scale OT problems. Its implementation also incorporates a rounding procedure to ensure the final solution satisfies the original OT constraint.

\cite{li2023fast} investigated the computational challenges associated with the optimal transport (OT) problem, particularly in large-scale applications. They aimed to develop a scalable first-order optimization algorithm that computes OT distances with \(\epsilon\)-approximation accuracy, addressing issues like inefficient error scaling and limited practical usability in existing methods. They introduced an entropy-regularized extra gradient method, leveraging bilinear minimax reformulation and adaptive learning rates. The method includes a novel adjustment mechanism to ensure stability and prevent undesirable solutions. The core contribution involves reformulating the OT problem with entropy regularization \eqref{eq:hreg} and solving it using extra gradient methods. They reformulated problem \eqref{eq:vich} as
\[
\min_{\{\mathbf{p}_i\}_{i=1}^{n}} \left\{\sum_{i=1}^{n} \va_i \langle \mathbf{c}_i,\mathbf{p}_i\rangle: \sum_{i=1}^{n} \va_i\mathbf{p}_i = \vb, \mathbf{p}_i\in\Delta_n\right\}
\]
where \(\mathbf{c}_i\) represents the \(i\)-th row of \(\mC\). To account for \(\mP\vone = \va\), they defined \(\{\mathbf{p}_i \in \Delta_n\}\) such that the \(i\)-th row of \(\mP\) is expressed as \(\va_i \mathbf{p}_i\). The authors examined the \(\ell_1\)-penalized variant of the above formulation, expressed as:
\[
\min_{\{\mathbf{p}_i\in\Delta_n\}_{i=1}^{n}} f_{\ell_1}\left(\{\mathbf{p}_i\}\right) = \frac{1}{2}\sum_{i=1}^{n} \va_i \langle \mathbf{c}_i,\mathbf{p}_i\rangle + \Vert\mC \Vert_{\infty} \left\Vert\sum_{i=1}^{n} \va_i\mathbf{p}_i - \vb\right\Vert_1
\]
This formulation resembles the one presented in \citep{jambulapati2019direct}, but with distinct definitions for the matrices. The authors assumed \(\Vert\mC\Vert_{\infty} = 1\) and introduced a set of auxiliary two-dimensional probability vectors \(\pmb{\mu}_j = [\mu_{j,+}, \mu_{j,-}] \in \Delta_2\), for \(1 \leq j \leq n\), which simplified the above problem to:
\[
\min_{\{\mathbf{p}_i\in\Delta_n\}_{i=1}^{n}} \max_{\{\pmb{\mu}_j\}_{j=1}^n} f \left(\{\mathbf{p}_i\},\{\pmb{\mu}_j\}\right) \xLeftrightarrow{\text{von Neumann’s minimax thm}} \max_{\{\pmb{\mu}_j\}_{j=1}^n} \min_{\{\mathbf{p}_i\in\Delta_n\}_{i=1}^{n}} f \left(\{\mathbf{p}_i\},\{\pmb{\mu}_j\}\right)
\]
where 
\[
f \left(\{\mathbf{p}_i\},\{\pmb{\mu}_j\}\right) = \frac{1}{2}\sum_{i=1}^{n} \va_i \langle \mathbf{c}_i,\mathbf{p}_i\rangle - \sum_{j=1}^{n} \mathbf{c}_j (\mu_{j,+} - \mu_{j,-})  + \sum_{i=1}^{n}\sum_{j=1}^{n} \va_i \mP_{ij} (\mu_{j,+} - \mu_{j,-}) 
\]
They augmented the objective function with entropy regularization terms as $f$ is neither strongly convex nor strongly concave.
\begin{align*}
    \max_{\{\pmb{\mu}_j\}_{j=1}^n} \min_{\{\mathbf{p}_i\in\Delta_n\}_{i=1}^{n}} F\left(\{\mathbf{p}_i\},\{\pmb{\mu}_j\}\right)&:= \\
    f \left(\{\mathbf{p}_i\},\{\pmb{\mu}_j\}\right) &+ \sum_{j=1}^n \tau_{\mu,j} H(\pmb{\mu}_j) - \sum_{j=1}^n \tau_{p,j} H(\mathbf{p}_j)
\end{align*}
where $H(\cdot)$ is the entropy, and $\{\tau_{\mu,j}\}$ and $\{\tau_{\mathbf{p},j}\}$ are a set of positive regularization parameters. In the proposed extra gradient method, in each iteration $t$, they update 
\begin{itemize}
    \item Updates w.r.t. $\{\mathbf{p}_i\}$: main sequence $\{\mathbf{p}_i^t\}$; midpoints $\{\Bar{\mathbf{p}}_i^t\}$
    \item Updates w.r.t. $\{\pmb{\mu}_j\}$: main sequence $\{\pmb{\mu}_j^t\}$; midpoints $\{\Bar{\pmb{\mu}}_j^t\}$; adjusted main sequence $\{\pmb{\mu}_j^{t,adjust}\}$
\end{itemize}
Using a single iteration of mirror descent, with the KL divergence selected to track the displacement:
\begin{align*}
    \pmb{\mu}_j^{next} &= \arg\max_{\pmb{\mu}_j\in\Delta_2} \left\{\left\langle \nabla_{\pmb{\mu}_j}F\left(\{ \mathbf{p}_i^{grad}\}_{i=1}^n,\{\pmb{\mu}_j^{grad}\}_{j=1}^n\right), \pmb{\mu}_j\right\rangle - \frac{1}{\eta_{\mu,j}}\text{KL}\left(\pmb{\mu}_j\Vert \pmb{\mu}_j^{current}\right)\right\} \\
    \mathbf{p}_i^{next} &= \arg\min_{\mathbf{p}_i\in\Delta_n} \left\{\left\langle \nabla_{\mathbf{p}_i}F\left(\{ \mathbf{p}_i^{grad}\}_{i=1}^n,\{\pmb{\mu}_j^{grad}\}_{j=1}^n\right), \mathbf{p}_i\right\rangle - \frac{1}{\eta_{p,i}}\text{KL}\left(\mathbf{p}_i\Vert \mathbf{p}_i^{current}\right)\right\}
\end{align*}
In each iteration $t$, the algorithm
\begin{enumerate}
    \item computes $\Bar{\pmb{\mu}}_j^{t+1}$ using $\pmb{\mu}_j^{t,adjust}$ and $\mathbf{p}_i^t$;
    \item updates $\Bar{\mathbf{p}}_i^{t+1}$ using $\pmb{\mu}_j^{t,adjust}$ and $\mathbf{p}_i^t$;
    \item updates $\pmb{\mu}_j^{t+1}$ using $\pmb{\mu}_j^{t,adjust}$ and $\Bar{\mathbf{p}}_i^{t+1}$;
    \item updates $\mathbf{p}_i^{t+1}$ using $\pmb{\mu}_j^{t+1}$ and $\mathbf{p}_i^t$;
    \item adjusts the current iterates by $\mu_{j,s}^{t+1,adjust} \propto \max\left\{\mu_{j,s}^{t+1}, e^{-B}\max\left\{\mu_{j,+}^{t+1}, \mu_{j,-}^{t+1} \right\} \right\}$, $s\in\{+,-\}$, and $B > 0$ is some parameter.
\end{enumerate}
Learning rates are adjusted based on marginal sums of the input distributions to ensure fast convergence. The proposed algorithm ensures convergence with guarantees to achieve an \(\epsilon\)-approximation, featuring an iteration complexity of \(\bigO(1/\epsilon)\) and a runtime of \(\bigO(n^2/\epsilon)\), aligning with the best-known theoretical results.

\subsubsection{Reduction-Based Methods}
\cite{blanchet2024towards} explored methods for obtaining $\epsilon$-approximate solutions to the OT problem with enhanced asymptotic efficiency, utilizing connections to traditional problems, including LP for packing and scaling of matrices. This paper seeks to provide nearly linear time algorithms that advance state-of-the-art performance while establishing theoretical connections with classical problems like bipartite matching. They demonstrated a reduction of the OT problem to maximum cardinality matching in bipartite graphs. This indicated that enhancing the runtime beyond $\bigO(n^2/\epsilon)$ would also lead to a more efficient algorithm for bipartite matching. This improvement has traditionally relied on advanced methods like fast matrix multiplication or dynamic graph algorithms.

As a type of linear optimization problem, a packing linear program is expressed as follows.
\[
Z^*=\max_{\vx\in\rr_+^n} \left\{\mathbf{c}^\top\vx: \mathbf{A}\vx \leq \mathbf{d}\right\}
\]
where $\mathbf{d}\in\rr^m_+$, $\mathbf{c}\in\rr^n_+$, and $\mathbf{A}\in\rr_+^{m\times n}$. The OT problem \eqref{eq:vich} reduces to a packing linear program by defining
\[
\Hat{\mC} = \Vert\mC\Vert_{\infty}\vone\vone^\top - \mC
\]
Then, $\Hat{\mC}$ is nonnegative, and replacing $\Hat{\mC}$ with $\mC$ and transforming the equality constraints to the inequality ones in \eqref{eq:vich}, the OT problem is reformulated as a packing problem. It can be solved in $\bigO(n^2/\epsilon)$.

Consider a non-negative matrix $\mathbf{A}$ and vectors $\mathbf{r},\mathbf{c}\in\rr_+^n$ that holds $\sum_{i=1}^n r_i = \sum_{i=1}^n c_i$, $\Vert\mathbf{A}\Vert_{\infty}\leq 1,\Vert\mathbf{r}\Vert_{\infty}\leq 1, \text{ and }\Vert\mathbf{c}\Vert_{\infty}\leq 1$. Let $\mathbf{X},\mathbf{Y}$ be diagonal and nonnegative matrices. These matrices are $(\mathbf{r},\mathbf{c})-$scale $\mathbf{A}$ if it holdas that $\mathbf{B} = \mathbf{X}\mathbf{A}\mathbf{Y}$, $\mathbf{B}\vone=\mathbf{c}$, and $\mathbf{B}^\top\vone=\mathbf{r}$. Alternatively, if $\Vert\mathbf{B}\vone-\mathbf{c}\Vert_1 + \Vert\mathbf{B}^\top\vone-\mathbf{r}\Vert_1 \leq \epsilon$, then $\mathbf{X},\mathbf{Y}$ are said to $\epsilon-$approximately $(\mathbf{r},\mathbf{c})-$scale $\mathbf{A}$. The problem of scaling a matrix involves finding matrices $\mathbf{X},\mathbf{Y}$ that $\epsilon-$approximately $(\mathbf{r},\mathbf{c})-$scale $\mathbf{A}$, assuming they exist. The dual of \eqref{eq:hreg} can be reformulated as a problem of scaling a matrix and solved in $\bigO(n^2/\epsilon)$.

\cite{quanrud2018approximating} introduced efficient algorithms for obtaining $\epsilon-$approximate solutions to the OT problem \eqref{eq:vich} by transforming it into more straightforward LP problems. The OT problem is reformulated into ``positive LPs,'' ``mixed packing and covering LPs,'' and ``packing LPs,'' leveraging their structure for efficient computation. The proposed algorithms focus on achieving $\epsilon-$approximations with significantly improved runtime complexity compared to existing methods. The reduction to positive LPs expresses OT as:
\[
\min_{\mP\in\rr^{n\times n}_+} \left\{ \langle \mC,\mP\rangle: \mP \va=\vb, \mP^\top\vone=\vone \right\}
\]
By leveraging relative approximation algorithms, the authors also reformulate OT into mixed packing and covering LPs:
\[
\min_{\mP\in\rr^{n\times n}_+} \left\{ \langle \mC,\mP\rangle: \mP \va\leq\vb, \mP^\top\vone\leq\vone \right\}
\]
and packing LPs:
\[
\min_{\mP\in\rr^{n\times n}_+} \left\{ \langle \vone,\mP\va\rangle: \mP \va\leq\vb, \Vert\mP\Vert_1\leq\lambda \right\}
\]
These formulations enable efficient computation of approximate solutions, particularly for large-scale problems with $\bigO(n^2/\epsilon)$ time complexity. To handle residual mass in approximate solutions, the authors introduce an ``oblivious transport" scheme that efficiently redistributes untransported mass.

\section{OT Problem Applications}\label{sec:applications}

OT problem has become a versatile tool across diverse applications in machine learning, data science, and beyond. Its ability to quantify distributional differences has found utility in generative models, domain adaptation, and clustering, among others. OT has also been employed in computer vision for image-to-image translation, point cloud alignment, feature matching, statistics for distribution shift detection, and Wasserstein barycenters. In addition to its typical applications, OT can also support decision-making in transportation and routing~\citep{MORADI2023109552,moradi2024two,MORADI2024110730,vehicles6030066,moradi2024electric,MORADI2025125183} and public health systems during pandemics~\citep{nazari2023}, as well as enhance population-based algorithms such as simulated annealing~\citep{moradi2022efficient}. Another notable application area is time-series data analysis, where OT provides a robust alternative to traditional methods like Dynamic Time Warping (DTW). This capability has shown to be particularly useful in domains such as cybersecurity, where aligning time-series data can aid in identifying system anomalies or understanding attack patterns. Similarly, in manufacturing systems, OT-based time-series alignment supports monitoring and optimizing industrial processes to ensure efficiency and resilience~\citep{Aftabi09122024,AFTABI2025125681}

A particularly innovative approach to time-series data analysis using OT is presented in~\citep{latorre2023otw}. The authors proposed Optimal Transport Warping (OTW) as an alternative to DTW, addressing its limitations in capturing time-series data's geometric and probabilistic structure. By framing the time series alignment as a transport problem, OTW improves upon traditional alignment methods by leveraging OT's inherent flexibility and geometric grounding. \cite{latorre2023otw} introduced a novel approach for aligning time-series data using the framework of OT. Traditional alignment methods like DTW often fail to capture the true geometric and probabilistic relationships within complex, high-dimensional time-series data. This work addresses these shortcomings by framing the alignment problem as an OT task, enabling a more flexible and robust alignment process.

The authors propose the OTW method, which models the alignment task as an OT problem between two time-series distributions. For two time-series \(\mathbf{x} = \{x_t\}_{t=1}^T\) and \(\mathbf{y} = \{y_t\}_{t=1}^T\), OTW seeks to compute a transport map \(\pi\) that minimizes the alignment cost:
\[
\min_{\pi \in \Pi(\mathbf{p}, \mathbf{q})} \sum_{i,j} \pi_{ij} c(x_i, y_j),
\]
where \(\Pi(\mathbf{p}, \mathbf{q})\) is the set of joint distributions with marginals \(\mathbf{p}\) and \(\mathbf{q}\), and \(c(x_i, y_j)\) is the cost of aligning \(x_i\) and \(y_j\). The transport cost matrix \(\mC\) is learned adaptively, ensuring alignment reflects temporal and feature-based similarities. The authors also incorporate temporal regularization, $R(\pi) = \sum_{i,j} \pi_{ij} \log(\pi_{ij})$, through an entropy term, balancing the need for temporal alignment and probabilistic structure. The algorithm employs Sinkhorn iterations for entropy-regularized OT to ensure computational efficiency and scalability for large datasets. This approach guarantees convergence to the optimal alignment while maintaining interpretability.

The proposed OTW method demonstrates significant advantages in aligning time series while respecting temporal dependencies and probabilistic distributions. By integrating OT principles into time series analysis, this work provides a robust alternative to traditional methods and establishes a foundation for future exploration in the field.

\section{Emerging Trends and Challenges in OT Problem}\label{sec:challenges}

The OT problem has become a cornerstone for numerous applications across machine learning, data science, economics, and beyond. However, the increasing complexity of modern datasets and demands for scalability and precision have revealed several emerging trends and unresolved challenges that shape the trajectory of this field. Computational advancements remain at the forefront of this evolution, with scalable algorithms designed for high-dimensional data becoming a pivotal area of focus.

A significant trend involves the integration of OT with machine learning frameworks. The synergy between deep learning and optimal transport has proven highly productive, finding use in areas such as GANs, cross-domain adaptation, and feature representation learning. Furthermore, graph-based OT methodologies unlock new potential in analyzing structured data, such as social networks and biological datasets. At the same time, federated learning presents opportunities for OT to enhance privacy-preserving computations. These trends exemplify OT's expanding role in complex, data-driven environments.

The expansion of OT problem variants presents another frontier of innovation. Unbalanced OT addresses distributions with unequal total mass, and dynamic OT, focusing on time-evolving distributions, has garnered substantial attention. Efficient computation of Wasserstein barycenters, critical for clustering and data aggregation tasks, continues to push the boundaries of feasible applications. However, theoretical guarantees for these emerging variants are often lacking, posing a challenge to their reliability and adoption.

Theoretically, the field faces persistent challenges in improving convergence guarantees for novel algorithms and establishing lower bounds on computational complexity. Robustness and stability issues remain a concern, particularly for methods operating in extreme parameter regimes. These theoretical gaps limit the reliability of existing methods and hinder their applicability in critical domains requiring precision and consistency.

Finally, as OT continues to penetrate diverse applications, societal implications and ethical concerns become increasingly relevant. The use of OT in fairness-aware algorithms highlights its potential to design equitable systems, but it also raises questions about interpretability and transparency. Moreover, as OT finds applications in sensitive areas such as surveillance and biased decision-making, ethical considerations must guide future research and deployment. Balancing the technical rigor of OT methods with their societal impact will be paramount to ensure responsible use.

\section{Conclusions}
\label{sec:conc}

OT has become a fundamental mathematical framework with profound implications across diverse domains. Its ability to quantify and align distributions has enabled groundbreaking advancements in machine learning, data science, computer vision, and numerous other fields. This report has explored the theoretical foundations of OT, cutting-edge computational algorithms, emerging challenges, and its applications in areas ranging from generative modeling to time-series analysis. The versatility of OT lies in its flexibility to adapt to different problem settings, such as unbalanced distributions, dynamic scenarios, and high-dimensional data. Advances in scalable algorithms and approximate methods have significantly expanded its applicability, making it feasible to address large-scale problems. However, challenges remain in achieving robust convergence, ensuring computational efficiency, and addressing the societal implications of OT's integration into real-world systems. As OT continues to evolve, its intersection with machine learning, graph theory, and optimization promises to unlock new opportunities for research and innovation. Addressing the unresolved theoretical and practical challenges will enhance the robustness and efficiency of OT algorithms and ensure their ethical and impactful deployment across diverse industries. By bridging theory and practice, OT solidifies its role as a modern computational and applied science cornerstone.

\bibliographystyle{plainnat}  
\bibliography{references}  %%% Remove comment to use the external .bib file (using bibtex).
%%% and comment out the ``thebibliography'' section.

%%% Comment out this section when you \bibliography{references} is enabled.
% \begin{thebibliography}{1}

% \bibitem{kour2014real}
% George Kour and Raid Saabne.
% \newblock Real-time segmentation of on-line handwritten arabic script.
% \newblock In {\em Frontiers in Handwriting Recognition (ICFHR), 2014 14th
%   International Conference on}, pages 417--422. IEEE, 2014.

% \bibitem{kour2014fast}
% George Kour and Raid Saabne.
% \newblock Fast classification of handwritten on-line arabic characters.
% \newblock In {\em Soft Computing and Pattern Recognition (SoCPaR), 2014 6th
%   International Conference of}, pages 312--318. IEEE, 2014.

% \bibitem{hadash2018estimate}
% Guy Hadash, Einat Kermany, Boaz Carmeli, Ofer Lavi, George Kour, and Alon
%   Jacovi.
% \newblock Estimate and replace: A novel approach to integrating deep neural
%   networks with existing applications.
% \newblock {\em arXiv preprint arXiv:1804.09028}, 2018.

% \end{thebibliography}

\end{document}